\documentstyle[prd,aps,epsfig,preprint]{revtex}
\begin{document}
\preprint{                                                 IASSNS-AST 96/21}
\draft
\title{\Large\bf        Matter-enhanced Three-flavor Oscillations         \\
                             and the Solar Neutrino Problem}
\author{                            G.~L.~Fogli                            }
\address{         Dipartimento di Fisica and Sezione INFN, Bari, Italy     }
\author{                             E.~Lisi                               }
\address{      Institute for Advanced Study, Princeton, New Jersey 08540
\\[-2mm]        and Dipartimento di Fisica and Sezione INFN, Bari, Italy   }
\author{                          D.~Montanino                             }
\address{         Dipartimento di Fisica and Sezione INFN, Bari, Italy     }
\maketitle
\begin{abstract}
We present a systematic analysis of the three-flavor 
Mikheyev-Smirnov-Wolfenstein (MSW) oscillation solutions to the solar 
neutrino problem, in the hypothesis that the two independent neutrino
square mass differences, $\delta m^2$ and $m^2$, are well separated: 
$\delta m^2 \ll m^2$. At zeroth order in $\delta m^2/m^2$, the relevant 
variables for solar neutrinos are $\delta m^2$ and two mixing angles, 
$\omega$ and $\phi$. We introduce new graphical representations of the 
parameter space $(\delta m^2,\,\omega,\,\phi)$, that prove useful both 
to analyze the properties of the electron-neutrino survival probability 
and to present the results of the analysis of solar neutrino data.
We make a detailed comparison between the theoretical predictions of the 
Bahcall--Pinsonneault standard solar model and the current experimental 
results on solar neutrino rates, and discuss thoroughly the MSW solutions 
found by spanning the whole three-flavor space
$(\delta m^2,\,\omega,\,\phi)$. The allowed regions can be radically 
different from the usual ``small mixing'' and ``large mixing'' solutions, 
characteristic of the usual two-generation MSW approach. We also discuss 
the link between these results and the  independent information on neutrino 
masses and mixings coming from accelerator and reactor oscillation searches.
\end{abstract}


\section{INTRODUCTION}

The measurement of the $Z\rightarrow \bar\nu \nu$ width at the CERN
Large Electron-Positron collider \cite{LEPc} has shown conclusively that 
there are three generations of light neutrinos. However, the problem as 
to whether the neutrinos have non-zero mass and mixing, as is the case 
for quarks, remains one of the most prominent experimental and theoretical 
problems in particle physics.

On the one hand, there is a vast and variegated set of experimental 
information, ranging from the negative  results of laboratory  neutrino 
oscillation searches at accelerators and reactors, to the possible 
indications for flavor transition processes in natural beams such as solar 
and atmospheric neutrinos. On the other hand,  current theoretical models 
(e.g., the see-saw mechanism \cite{Sees})  allow many different textures 
for the three-generation neutrino mass matrix. Thus, it seems appropriate, 
at this stage, to keep to a minimum the number of {\em a priori\/} 
assumptions in the  interpretation of the experimental results.

In this work, we focus on the three-flavor Mikheyev-Smirnov-Wolfenstein
(MSW) \cite{Wo78} neutrino oscillation solutions to the  solar neutrino 
problem \cite{Ba89}. The only hypothesis that we make is to assume a 
neutrino mass spectrum as in Fig.~1.  More precisely, we assume  one of 
the two independent square mass differences, $m^2$, to be very large 
(more than one order of magnitude higher) as compared to the smallest one, 
$\delta m^2$, which drives the matter-enhanced oscillations. 

At zeroth order in $\delta m^2/m^2$, the relevant MSW parameters
(see Sec.~III) are $\delta m^2$ and two mixing angles, $\omega$ and $\phi$,
ranging in the first quadrant $[0,\,\pi/2]$. We introduce new graphical
representations of the parameter space $(\delta m^2,\,\omega,\,\phi)$,
that prove useful in analyzing the properties of the three-flavor MSW
probability, as well as in showing the detailed results of fits to the
experimental data. The parameter space is studied exhaustively, and
all the solutions are discussed thoroughly. We find three-flavor solutions 
to the solar neutrino problem that are considerably different from 
the usual two-flavor results. 

The present study is more complete and detailed
than previous three-flavor phenomenological analyses of the solar neutrino 
problem  performed by other authors \cite{Mi88,Sh92,Ha93,Jo93,Ma95} and by 
ourselves \cite{Fo94,Ve95,Mo95}. The work presented in this paper is part 
of a wider research program, in which we intend to analyze the world neutrino 
oscillation data under the sole hypothesis represented by the spectrum 
in Fig.~1. The analysis of the most constraining accelerator and reactor
experiments has already been completed \cite{Li95}. We discuss the link 
between the information obtained in the present work from solar neutrinos 
and the results obtained in \cite{Li95} from laboratory oscillation searches. 
Concerning atmospheric neutrinos, a complete three-flavor analysis  
(partially addressed in \cite{Fo94,Ve95,Mo95,Li95,FL95}) is in progress.

This paper has the following structure. In Sec.~II we review the current 
theoretical expectations and experimental observations of the solar neutrino 
fluxes, and show for completeness the usual two-generation MSW fits to 
the data. In Sec.~III we present the three-flavor oscillation formalism 
and show the new graphical representations of the parameter space.
The properties of the $\nu_e$ survival probability are also discussed. 
In Sec.~IV we explore the parameter space exhaustively, and find the 
three-flavor solutions to the solar neutrino problem. Both in Sec.~III 
and in Sec.~IV we show how  solar neutrino results can be compared
to the  information coming from accelerator and reactor neutrino experiments. 
We conclude and summarize our work in Sec.~V.

\section{SOLAR  NEUTRINO  PROBLEM  AND  TWO-GENERATION  MSW  SOLUTIONS}

In this section we present the experimental data used in this work,
and compare them with the theoretical expectations. Particular attention
is paid to correlation effects. Although our paper is focused on the 
three-flavor MSW mechanism, we also show, for the sake of completeness, 
our fit to the data in the simpler and more familiar two-generation approach.

\subsection{Experimental data and theoretical predictions}

There are four operating solar neutrino experiments: The Homestake 
\cite{Da94} chlorine (Cl) detector,  the GALLEX \cite{GALL} and SAGE 
\cite{SAGE} gallium (Ga) detectors, and the Kamiokande \cite{Kami} 
water-Cherenkov detector. The observed solar neutrino rates are shown 
in Table~I, along with the corresponding theoretical values as predicted 
in the recent refined Standard Solar Model (SSM) by Bahcall and Pinsonneault 
(BP) \cite{Ba95}.  The significant deficit of observed neutrinos 
constitutes the well-known ``solar neutrino problem'' \cite{Ba89}.

The information contained in Table~I does not characterize completely
the significance of the solar neutrino problem, since the theoretical
uncertainties (last column of Table~I) are not independent. It has 
been shown in a previous work  \cite{Fo95} that the correlation of 
the uncertainties can be calculated  analytically, starting from the 
errors affecting the SSM input parameters  and the  neutrino capture 
cross-sections. 

In Fig.~2 we show  the  experimental and theoretical 
$99\%$  C.L.\ contours  ($\Delta\chi^2=9.21$, $N_{DF}=2$)  for any  
couple of experiments  (the GALLEX and SAGE data have been combined 
quadratically). The projections of  the ellipses onto the coordinate 
axes represent, with good approximation,  the $\pm 3\sigma$ errors for 
each experiment separately. Notice the strong correlation of the 
theoretical errors. If the correlations were neglected, the theoretically 
allowed regions would be larger and the  discrepancy with the experiments 
would be underestimated 
\cite{Fo95,FL94,Ha94,Ga94}.

The last relevant piece of data is the night-day $(N-D)$ asymmetry of the 
solar neutrino flux as measured in the Kamiokande detector \cite{Ta95,Kami}:
\begin{equation}
\frac{N-D}{N+D} = 0.07\pm0.07 ({\rm stat}) \pm0.04 ({\rm syst}) \quad.
\end{equation}
The above value is consistent with no asymmetry (i.e., no oscillations),
and thus contributes to exclude a part of the neutrino oscillation 
parameter space.

\subsection{Two-generation MSW analysis}

The MSW mechanism \cite{Wo78} of matter-enhanced neutrino oscillations is
one of the most promising candidates for the solution of the solar neutrino
problem (for reviews, see \cite{Bi87,Mi89,Ku89,Hx95}). In the hypothesis 
that only two neutrino generations are involved in the $\nu_e$ transitions, 
the relevant parameters are one mass square difference, $\delta m^2$, and 
one mixing angle, $\omega$. 

Even in the simple two-family scenario, the calculation of the electron 
neutrino survival probability, $P(\nu_e\rightarrow\nu_e)$, as well as the 
data analysis, are not trivial tasks and require refined computer codes. 
We take this opportunity to mention a few state-of-the-art features of 
our codes. We smear $P(\nu_e\rightarrow\nu_e)$ over the neutrino production 
region in the most general way, i.e., we also consider neutrino production 
points off the line that joins the center of the sun and the detector. This
requires a double (radial and azimuthal) space integration in the sun, 
as well as the evaluation  of the solar electron density gradient along 
off-radial trajectories (instead of the simple derivative). The possibility 
of a double resonance (see, e.g., \cite{Ku89,Ro86}) is included. We also 
include the (computer time-consuming) earth regeneration effect \cite{We87} 
at each detector location, and in particular we calculate the night-day 
asymmetry \cite{Hi87} at the Kamiokande site. The earth electron density 
is modeled as a step function with five steps \cite{Fo94}, corresponding 
to the relevant radial shells \cite{An89}. The theoretical uncertainties
and their correlation effects are taken into account in the statistical
$\chi^2$-analysis of the data \cite{Fo95}.

The results of our two-flavor analysis are shown in Fig.~3. The thick,
solid  lines represent the contours of the regions allowed at $95\%$ C.L.\
($\Delta\chi^2=5.99$, $N_{DF}=2$). In the first four panels, dotted lines 
represent curves of iso-signal. In the last two panels, all data are
combined without and with theoretical errors, and the allowed regions
at $90\%$, $95\%$ and  $99\%$ C.L.\ are shown. Notice the widening of the 
``small'' and ``large'' mixing angle solutions when the theoretical
uncertainties are taken into account. A third solution appears at $99\%$
C.L.\ for large angles and small $\delta m^2$.

When all the uncertainties are included, the $\chi^2$ function reaches 
its minimum, $\chi^2_{\rm min}=0.74$, at $\sin^2 2\omega=8.06\times10^{-3}$
and $\delta m^2=5.21\times10^{-6}$ eV$^2$. The secondary (large angle) 
minimum is reached at $\sin^2 2\omega=0.64$ and 
$\delta m^2=1.45\times10^{-5}$ eV$^2$, with a value 
$\chi^{2\prime}_{\rm min}=1.89$---not a bad fit.

The MSW solutions in Fig.~3  compare well with the most recent 
two-generation analyses of the solar neutrino problem
\cite{Ha94,Ga94,Bi94,Fi94,Kr94,Pe95,Ha95,Kr95}, 
modulo small differences due to the input data and their treatment. 
In particular, we have verified that, excluding the earth regeneration 
effect, our results become very similar to those obtained very recently 
by Bahcall and Krastev in \cite{Kr95}, where the same solar model and 
experimental data were used as input.

\section{THREE-FLAVOR  MSW  SOLAR  NEUTRINO  OSCILLATIONS: 
                      THEORETICAL  ASPECTS}

In this Section we recall the formalism that will be used in the 
three-flavor analysis, under the hypothesis that the neutrino mass
spectrum is as in Fig.~1. We introduce new graphical representations
of the  parameter space, and use them  to discuss the properties
of the three-flavor MSW probability $P(\nu_e\rightarrow\nu_e)$.

\subsection{A simple three-flavor framework and its parameter space}

In the general three-generation case, the flavor eigenstates
$\nu_{\alpha}$ $(\alpha=e,\,\mu,\,\tau)$ are a superposition of three 
mass eigenstates $\nu_{i}$ $(i=1,\,2,\,3)$:
$\nu_\alpha = U_{\alpha i} \nu_i$. The (unitary) neutrino  mixing matrix 
$U_{\alpha i}$ is usually parameterized in the same way as the standard 
Cabibbo-Kobayashi-Maskawa (CKM) mixing matrix in the quark sector 
\cite{PDGR}, involving three mixing angles, 
$(\theta_{12},\,\theta_{23},\,\theta_{13})\in [0,\,\pi/2]$, and one
CP-violating phase, $\delta$. Redefining the angles as
$\omega\equiv\theta_{12}$, 
$\psi\equiv\theta_{23}$, and
$\phi\equiv\theta_{13}$, the standard parameterization reads:
\begin{equation}
\footnotesize
\left(\begin{array}{c}\nu_e\\ \nu_\mu\\ \nu_\tau\end{array}\right)= 
\left(\begin{array}{ccc} 					
\cos\omega\cos\phi &							
        \sin\omega\cos\phi & 						
             \sin\phi e^{-i\delta}\\					
-\sin\omega\cos\psi-\cos\omega\sin\psi\sin\phi e^{i\delta} &		
        \cos\omega\cos\psi-\sin\omega\sin\psi\sin\phi e^{i\delta} &     
             \sin\psi\cos\phi \\					
\sin\omega\sin\psi-\cos\omega\cos\psi\sin\phi e^{i\delta} & 		
         -\cos\omega\sin\psi-\sin\omega\cos\psi\sin\phi e^{i\delta} &	
              \cos\psi\cos\phi						
 \end{array}\right)
\left(\begin{array}{c}\nu_1\\ \nu_2\\ \nu_3\end{array}\right).
\end{equation}

Permutations of the mass eigenstate labels $(1,\,2,\,3)$ do not change
the physics, provided that the mixing angles $(\omega,\,\psi,\,\phi)$ are
left to range in the first quadrant \cite{PDGR}. The parameter set is 
completed by two independent neutrino square mass differences, that we 
write as:
\begin{equation}
\delta m^2 \equiv m^2_2 - m^2_1\quad,\quad m^2\equiv m^2_3-m^2_2\quad,
\end{equation}
so that the general three-flavor parameter space is  
$(\delta m^2,\,m^2,\,\omega,\,\psi,\,\phi,\,\delta)$.

In principle, a complete phenomenological analysis of three-flavor neutrino 
oscillations should span the entire 6-dimensional manifold 
$(\delta m^2,\,m^2,\,\omega,\,\psi,\,\phi,\,\delta)$. This task would be 
exceedingly difficult---and perhaps not really useful---at this time, 
due to vastness  of the parameter space, as opposed to the scarcity of 
evidences in favor of neutrino oscillations.

A much more manageable framework is obtained under  the simple hypothesis
that the mass spectrum has the property shown in Fig.~1: 
$\delta m^2 \ll m^2$.  The two cases (a) and (b) in Fig.~1 lead to the 
same solar neutrino physics. We will always refer to (a) in the following, 
and conventionally label the mass eigenstates in the order of increasing 
mass $(m_1<m_2<m_3)$. In this case, as far as solar neutrinos are concerned,
the two quasi-degenerate mass eigenstates $\nu_1$ and $\nu_2$
participate actively to matter-enhanced transitions, while the ``lone''
state $\nu_3$ is a spectator neutrino that reveals its presence only 
through the mixing \cite{Ku86,To87,Ki87}.

At zeroth order in $\delta m^2/m^2$, the following properties hold
(see \cite{Fo94,Li95} and references therein):
    (1) solar neutrino experiments probe only the subspace
$(\delta m^2,\,\omega,\,\phi)$; 
    (2) terrestrial (i.e., accelerator, reactor, atmospheric)  
neutrino experiments probe only the subspace $(m^2,\,\psi,\,\phi)$; 
    (3) effects related to the CP-violating phase $\delta$ are unobservable,
so that the mixing matrix in Eq.~(2) can be taken as real for our purposes.

In other words,  solar neutrino experiments explore the space 
$(\delta m^2,\,U^2_{e1},\,U^2_{e2},\,U^2_{e3})$ that is the small mass 
square difference and the three matrix elements related to $\nu_e$ 
(the fast $m^2$-driven oscillations being  effectively averaged out) 
\cite{Fo94}. Terrestrial (accelerator, reactor, atmospheric) neutrino
experiments explore instead the space 
$(m^2,\,U^2_{e3},\,U^2_{\mu3},\,U^2_{\tau3})$ that is the large mass square 
difference and the three matrix elements related to $\nu_3$ 
(the slow $\delta m^2$-driven oscillations being effectively frozen) 
\cite{Li95}.

The five mixing matrix elements $U_{ei}$ and $U_{\alpha3}$ probed by solar 
and terrestrial oscillation searches satisfy the two unitarity conditions:
\begin{mathletters}
\begin{equation}
		U^2_{e3}+U^2_{\mu3}+U^2_{\tau3}	= 1	\quad,
\end{equation}
\begin{equation}
		U^2_{e1}+U^2_{e2}+U^2_{e3}	= 1	\quad.
\end{equation}
\end{mathletters}

In \cite{Li95} it has been shown that Eq.~(4a) can be embedded in a 
triangular representation of the space spanned by the the ``lone'' 
neutrino state: $\nu_3=U_{e3}\,\nu_e+U_{\mu3}\,\nu_\mu+U_{\tau3}\,\nu_\tau$.
In Fig.~4, we embed analogously the second unitarity condition, Eq.~(4b),
in a triangular representation of the space spanned by the electron
neutrino state: $\nu_e=U_{e1}\,\nu_1+U_{e2}\,\nu_2+U_{e3}\,\nu_3$. 
The triangle in Fig.~4 has equal sides, unit height, and corners 
corresponding to the mass eigenstates $\nu_1$, $\nu_2$, $\nu_3$. A generic 
state $\nu_e$ is represented by a point in the triangle. The $U^2_{ei}$ 
are identified with the heights projected from $\nu_e$. The sum of such 
heights is always equal to the total (unit) height of the triangle.
In Fig.~4 we also chart, in the lower subplot, the triangle coordinates 
in terms of $\omega$ and $\phi$. The usual  two-generation limit is
reached for  $\phi=0$, that is for $\nu_e$ on the side joining 
$\nu_1$ to $\nu_2$ (and thus decoupled from $\nu_3$).

In Fig.~5 we show a synoptic presentation of the two triangular graphs 
introduced in Fig.~4 of this work and in Fig.~2 of \cite{Li95}. 
On the left (right) side we display the parameterization of solar 
(terrestrial) neutrino oscillations, including the expansion of the 
relevant mixing matrix elements in terms of the mixing angles. Note that 
both  solar  and terrestrial neutrinos probe the element $U_{e3}$, i.e., 
the angle $\phi$, as discussed in \cite{Fo94,Ve95,Mo95,Li95} and also 
in \cite{Mi95,Bi95}. This important link  between   the 
experiments sensitive to $\delta m^2$ (solar) and those sensitive to 
$m^2$ (terrestrial) will be utilized in  Sec.~IV.

\subsection{Electron neutrino survival probability}

The electron neutrino survival probability, $P(\nu_e\rightarrow\nu_e)$,
is the fundamental quantity to compute in the solar neutrino analysis.  
In this section we study some properties of the three-flavor MSW survival 
probability, $P_{3\nu}^{\rm MSW}$, as a function of $\omega$, $\phi$, 
and $\delta m^2/E_\nu$  ($E_\nu$ being the neutrino energy). We start 
with a brief review of a well-known analytical approximation to 
$P_{3\nu}^{\rm MSW}$, that has been used in the present work. In order
to avoid unnecessary complications, in this section and in the related
figures we ignore temporarily the  smearing of $P_{3\nu}^{\rm MSW}$
over the neutrino production region, and the earth regeneration
effect. These effects will be  included, however, in the global analysis 
of Sec.~IV.

In the limit $\delta m^2 \ll m^2$,  the probability $P_{3\nu}^{\rm MSW}$ 
takes the simple form \cite{Ku86,Mi88}:
\begin{equation}
 P_{3\nu}^{\rm MSW}= \cos^4\phi\, P_{2\nu}^{\rm MSW}+\sin^4\phi \quad,
\end{equation}
where  $P_{2\nu}^{\rm MSW}$ is the MSW probability in the 
two-generation  limit, provided that the solar electron  density  $N_e(x)$ 
at any point $x$ is effectively replaced by $N_e(x)\cos^2\phi$.

A very accurate and widely used analytical approximation to 
 $P_{2\nu}^{\rm MSW}$ is the so-called ``Parke's formula'' \cite{Pa86}:
\begin{equation}
P_{2\nu}^{\rm MSW} = \frac{1}{2}+\left(\frac{1}{2}-\Theta\,P_C\right)
\cos2\omega\cos2\omega^0_m \quad.
\end{equation}
This approximation was also studied by other authors \cite{Ha86}.
In Eq.~(6), $P_C$ is the Landau-Zener-Stueckelberg crossing probability 
\cite{La32}, that we use in the  improved form \cite{Pe88,Kr88} valid for a 
close-to-exponential radial density; $\omega^0_m$ is the mixing angle in 
matter at the production point. The step function $\Theta$ \cite{Ku89} 
switches from 0 (non-resonant propagation case) to 1 (resonant propagation 
case) if the mixing angle $\omega$ in matter can assume the value $\pi/4$ 
at a point $x$ along the neutrino trajectory in the sun:
\begin{equation}
\omega_m(x)=\frac{\pi}{4} \quad
\Longleftrightarrow\quad\delta m^2\, 
\frac{\cos2\omega}{\cos^2\phi} = 2\sqrt{2} G_F N_e(x)
E_\nu \quad.
\end{equation}
Equation~(7) numerically  reads
\begin{equation}
\frac{\delta m^2}{{\rm eV}^2}\,
\frac{\cos2\omega}{\cos^2\phi} = 1.60\times10^{-5}\,\frac{E_\nu}{\rm MeV}
\frac{N_e(x)}{N_e(0)}\quad,
\end{equation}
where $N_e(0)$ is the electron density at the center of the sun \cite{Ba95}. 

A technical remark is in order. In the literature, the numerical validity 
of the approximations implicit in Eq.~(6) has been discussed \cite{Kr88,Br95}
mainly for $\omega$ (and $\phi$ \cite{Ku89})  in the first octant 
$[0,\,\pi/4]$. Since we are interested in analyzing the full mixing angle 
space (the first quadrant), we have calculated the function 
$P_{3\nu}^{\rm MSW}=P_{3\nu}^{\rm MSW}(\delta m^2/E_\nu)$
with a Runge-Kutta integration of the three-flavor MSW equations,
for a number of representative $(\omega,\,\phi)$ values in $[0,\,\pi/2]$.
The agreement between the analytical approximation of Eqs.~(5,6)
and the ``exact'' (but much more time-consuming) Runge-Kutta calculation is
as good in the second octant as it is in the first, provided that 
$\delta m^2/E_\nu$ is taken above $\sim 10^{-8}$ eV$^2$/MeV . For  
values of $\delta m^2/E_\nu$ lower than $\sim 10^{-8}$ eV$^2$/MeV, one 
should more properly consider just-so vacuum oscillations \cite{Gl87}.

There are a few interesting limits for $P_{3\nu}^{\rm MSW}$. If
$\delta m^2/E_\nu\gg 10^{-5}$ eV$^2$/MeV, the MSW mechanism is not
effective and $P_{3\nu}^{\rm MSW}$ tends to its vacuum value 
$P_{3\nu}^{\rm vac}$ given by
\begin{eqnarray}
P_{3\nu}^{\rm vac}&=& \cos^4\phi\, P_{2\nu}^{\rm vac}+ \sin^4\phi
                                                                  \nonumber\\
                  &=& \cos^4\phi\left(1-\frac{1}{2}\sin^2 2\omega\right)
                                                               + \sin^4\phi\\
                  &=& 1-2(U^2_{e1}U^2_{e2}+U^2_{e2}U^2_{e3}+U^2_{e3}U^2_{e1})
                                                             \quad,\nonumber
\end{eqnarray}
where the fast $(\delta m^2/E_\nu)$-driven oscillations have been averaged
to $1/2$.

Given the above expression for $P_{3\nu}^{\rm vac}$, and Eqs.~(5,6) for
$P_{3\nu}^{\rm MSW}$, it follows that $P_{3\nu}^{\rm MSW}=P_{3\nu}^{\rm vac}$,
for arbitrary values of $(\delta m^2/E_\nu)$, 
in (at least) three cases:
\begin{mathletters}
\begin{eqnarray}
            \omega=0             & \quad \longrightarrow \qquad & 
P_{3\nu}^{\rm MSW}=  \;\,\cos^4\phi+\sin^4\phi\;\, =P_{3\nu}^{\rm vac}\quad,\\
            \omega=\frac{\pi}{4} & \quad \longrightarrow \qquad &
P_{3\nu}^{\rm MSW}=\frac{1}{2}\cos^4\phi+\sin^4\phi=P_{3\nu}^{\rm vac}\quad,\\
            \omega=\frac{\pi}{2} & \quad \longrightarrow \qquad & 
P_{3\nu}^{\rm MSW}=  \;\,\cos^4\phi+\sin^4\phi\;\, =P_{3\nu}^{\rm vac}\quad. 
\end{eqnarray}
\end{mathletters}
The above properties are exact, i.e., they do not depend on the approximations
implicit in Eq.~(6). A subcase of Eq.~(10b) is obtained in the ``maximal
mixing scenario'' \cite{Max1}, corresponding to the center of the triangle
in Fig.~4 ($\sin^2\omega=1/2$, $\sin^2\phi=1/3$).

A further property follows from Eqs.~(5,6,9):
\begin{equation}
P_{3\nu}^{\rm MSW} \Bigl|_{\omega^0_m=\frac{\pi}{4}}\; = 
                                  \;\frac{1}{2}\cos^4\phi + \sin^4\phi\; = \;
P_{3\nu}^{\rm vac} \Bigl|_{\omega=\frac{\pi}{4}}\quad.
\end{equation}
This property, however, is not exact and has the same limits of validity of 
Eq.~(6) (see also the discussion at the end of this section).

The equations (9--11) are helpful in understanding the behavior of 
$P_{3\nu}^{\rm MSW}$ as a function of its arguments $(\delta m^2/E_\nu,\,
\omega,\phi)$, as we discuss now in some detail. 

In Fig.~6, curves of constant $P_{3\nu}^{\rm MSW}$ are shown for selected 
values of $\delta m^2/E_\nu$ in the triangular  representation. In the first 
subplot ($\delta m^2/E_\nu\gg 10^{-5}$ eV$^2$/MeV), it is 
$P_{3\nu}^{\rm MSW}=P_{3\nu}^{\rm vac}$ in the whole triangle, 
and the iso-probability curves are circles [as it is easily derived from
Eq.~(9)], with a minimum value $P_{3\nu}=1/3$ at the triangle center. 
In the second subplot ($\delta m^2/E_\nu = 1.75\times10^{-5}$ eV$^2$/MeV), 
the curves  begin to be deformed by the MSW effect for $\nu_e$ close to 
$\nu_1$ (the low left corner). Notice, however, that just along the lines
at $\omega=0,\,\pi/4,\,\pi/2$ (refer to Fig.~4 also) the probability  takes 
the same values as in the previous plot 
($P_{3\nu}^{\rm MSW}=P_{3\nu}^{\rm vac}$), as expected from Eqs.~(10a--c). 
These properties also hold for the remaining four subplots where,
however, the increasingly important MSW effect prevents a  graphical 
resolution of the iso-lines at small $\omega$ values, i.e., close
to the left side of the triangle.

In Fig.~7 we map the triangular parameter space onto a ``square''
bilogarithmic plot with coordinates $(\tan^2\omega,\tan^2\phi)$.
This new representation has the advantage that the ``triangle corners''
are infinitely expanded, although the nice symmetry  properties of 
the triangular representation are lost. A similar representation 
was introduced in \cite{Li95} to map the  parameter
space of terrestrial neutrinos.  The values of $\delta m^2/E_\nu$ 
and the iso-probability curves in Fig.~7 correspond exactly to those in 
Fig.~6. In Fig.~7 we also draw two curves (thick, dashed lines) 
corresponding to $\omega^0_m=\pi/4$ [Eq.~(7) with $x=0$] and $\omega=\pi/4$.
These curves separate the zones where the propagation is non-resonant
from those where it is resonant, as indicated in all panels.
The behavior of $P_{3\nu}^{\rm MSW}$ is most interesting near the curve
at $\omega^0_m=\pi/4$ and in the resonance region. Notice,
for the subplots at $\delta m^2/E_\nu=1.45\times 10^{-5}$ eV$^2$/MeV
or lower, the appearance of a depletion zone for $P_{3\nu}^{\rm MSW}$
at small values of $\omega$ (the ``small angle MSW solution'').

In Fig.~7, the properties expressed by Eqs.~(10a-b) can be easily  checked. 
The implications of Eq.~(11) are instead more subtle; if one takes  any two 
points at the same  height (same $\phi$) on the dashed curves 
($\omega=\pi/4$ and  $\omega^0_m=\pi/4$), then the probabilities $P_{3\nu}$ 
at these conjugate points are equal. It may be said, in a figurative way,
that the MSW mechanism ``cuts'' the $(\omega,\,\phi)$ plane along the 
dashed lines at $\omega=\pi/4$ and $\omega^0_m=\pi/4$, ``separates''
the edges (that would coincide for  vacuum oscillations) and ``fills'' 
the cut zone with a new resonant region where the survival probability can 
be much lower than in the vacuum case. The  probabilities in the ``old'' 
regions at the right and left of the two dashed lines are only weakly 
(adiabatically) different from the corresponding vacuum oscillation values.

We finally show in Fig.~8  the behavior of $P_{3\nu}^{\rm MSW}$  for six 
representative values of $\phi$, in a bilogarithmic 
$(\delta m^2/E_\nu,\,\tan^2\omega)$  mass-mixing plane (a perhaps more 
familiar representation). As in the previous figure, the two thick, 
dashed  lines correspond to the curves at $\omega=\pi/4$ and
 $\omega^0_m=\pi/4$, that separate non-resonant and resonant
propagation regions. The curves of iso-probability in the resonant region 
have the characteristic triangular shape \cite{Mi89}.  Notice that the 
variations of $P_{3\nu}^{\rm MSW}$ within the mass-mixing plane decrease 
as $\phi$ increases and the MSW effect is suppressed.

A final remark about the calculation of the survival probability.
As shown in \cite{Kr88}, Eq.~(6) is a good approximation except for 
the particular case of a  neutrino created close to the resonance 
$(\omega^0_m\simeq\pi/4)$ {\em and\/} with very small vacuum mixing 
($\tan^2 \omega \lesssim 10^{-4}$). This situation would correspond to a 
very narrow strip, localized along the thick, dashed curves in Figs.~7 
and~8 for $\tan^2 \omega \lesssim 10^{-4}$. The local discrepancy with the 
exact probability is, however, unimportant in the calculation of the neutrino 
rates, being effectively suppressed by the integration over the neutrino 
production  region and energy spectrum, and being confined to values of 
$\omega$ where no solution to the solar neutrino problem is found.

\section{THREE-FLAVOR  MSW  SOLAR  NEUTRINO  OSCILLATIONS:
                       PHENOMENOLOGY}

In this section we present the results of our global three-flavor MSW 
analysis of solar neutrino data (Table~I) in the parameter space 
$(\delta m^2,\,\omega,\,\phi)$. We span the coordinates\ 
$\delta m^2\otimes \tan^2\omega\otimes \tan^2\phi$\ in the range\
$[10^{-7}{\rm \ eV}^2,\,10^{-3}{\rm \ eV}^2]\otimes[10^{-6},\,10^{-2}]
\otimes [10^{-4},\,10^{-4}]$\ and analyze the data through a $\chi^2$ 
statistic, including experimental and theoretical errors and their
correlations. 

We find the absolute minimum, $\chi^2_{\rm min}=0.74$, at
$\delta m^2 = 5.21\times 10^{-6}$ eV$^2$, $\omega=2.57^{\circ}$,
$\phi = 0^{\circ}$. It coincides with the minimum of the two-generation
analysis (Sec.~II~B). The surfaces at $\chi^2=\chi^2_{\rm min}+\Delta\chi^2$,
with $\Delta\chi^2=6.25$, 7.82, and 11.34, define the regions allowed
at $90\%$, $95\%$ and $99\%$ C.L.\ for three degrees of freedom.
We present representative sections of these three-dimensional allowed regions
in the bilogarithmic mass-mixing plane $(\delta m^2,\,\tan^2\omega)$
at several values of $\tan^2\phi$, and in the bilogarithmic
mixing-mixing plane $(\tan^2\omega,\tan^2\phi)$ at several values of
$\delta m^2$. We also show representative fits to separate pieces of data.
We finally discuss how this solar neutrino analysis is linked to the
terrestrial neutrino oscillation constraints obtained in \cite{Li95}.

\subsection{Analysis in the mass-mixing plane}

In Fig.~9 we show the sections of the allowed regions in the bilogarithmic
plane $(\tan^2\omega,\,\delta m^2)$, for representative values of 
$\tan^2\phi$. Notice than both $\omega$ and $\phi$ are allowed to range
above the first octant in the fit. In the first panel, $\tan^2\phi$
is zero and  purely  two-generation oscillations take place. 
In fact, this subplot is a mapping of Fig.~3, modulo the different number
of degrees of freedom in the definition of the C.L.\ contours.
A moderate increase of $\tan^2\phi$ from 0.1 up to 0.4
produces an increasing deformation of the allowed contours. The contours 
tend to merge in the upper part, corresponding to the upper side of the 
MSW ``triangles'' of Fig.~8.  The two separate solutions merge in a 
single connected solution at $\tan^2\phi\simeq 0.5$, as also shown in 
\cite{Ha93,Ma95}. For increasing value of $\tan^2\phi$, the allowed region
becomes  broader and less structured, and then rapidly shrinks
and disappears at $\tan^2\phi\gtrsim1.5$. Figure~8 helps in
understanding the flattening of the allowed region in Fig.~9 at
the higher values of $\tan^2\phi$. 

From Fig.~9 it can be noticed that, although there are no solutions for 
$\omega>\pi/4$, the fit can  still be acceptable for $\phi>\pi/4$. In 
Fig.~10 we then show the values of $\chi^2-\chi^2_{\rm min}$ as a function 
of $\tan^2 \phi$, for $\delta m^2$ and $\tan^2\omega$  unconstrained. 
The minimum is reached at $\phi=0^{\circ}$, meaning that the two-generation 
limit is preferred by the present data. However, three-flavor solutions are 
allowed for values of $\tan^2\phi$ as high as 1.4 (i.e., 
$\phi\simeq 50^{\circ}$) at $90 \%$ C.L. There are no acceptable 
solutions at $99\%$ C.L.\ for $\tan^2\phi\gtrsim2$.

The behavior of the solutions in Fig.~9 can be understood better 
by separating the information coming from single pieces of data.
In Fig.~11 we show the separate fits to the gallium, chlorine and
Kamiokande (total rate and night-day asymmetry) measurements,
in the same coordinates as in Fig.~9, for three values of $\tan^2\phi$.
The thick, solid line represents contours at $95\%$ C.L. The allowed regions
are marked by stars. The dotted lines represent contours of iso-signal.
The structured shape of the iso-signal curves is globally similar to
the iso-probability curves in Fig.~8. The regions allowed by the separate
pieces of data become wider for increasing $\phi$, and even extend 
above $\omega=\pi/4$, but their mutual compatibility do not necessarily
increase. In particular, notice in Fig.~11  that the chlorine data play a 
major role in confining $\omega$ below $\pi/4$ in the global fit of Fig.~9.

We have found three-flavor solutions to the solar neutrino problem at 
relatively large $\phi$. These solutions   are characterized by
a survival probability, $P_{3\nu}^{\rm MSW}$, that varies more slowly 
with energy as compared to the two-generation case (see Fig.~8). Probing 
the large-$\phi$ three-flavor solutions will therefore represent
a formidable challenge to the new generation of solar neutrino
experiments, SuperKamiokande \cite{SKam}, the  Sudbury Neutrino Observatory 
(SNO) \cite{Sudb}, and the Imaging of Cosmic and Rare Underground Signals
(ICARUS) \cite{Icar},   that are meant to probe the energy-dependence
of the solar neutrino flux at earth.

\subsection{Analysis in the mixing-mixing plane}

In Fig.~12 we show the sections of the allowed region in the mixing-mixing
plane $(\tan^2\omega,\,\tan^2\phi)$ for representative values of 
$\delta m^2$. In all panels, the two-generation limit is reached
for $\tan^2\phi\rightarrow 0$ (lower side), and the solutions lie along 
one or more of a three-sided {\large $\sqcap$}-shaped band. The  sides 
correspond to ``small $\omega$,'' ``large $\omega$,'' and ``large $\phi$'' 
solutions. The ``small $\omega$'' and ``large $\omega$'' solutions protrude
in the three-flavor space from the usual MSW solutions in the two-generation
limit. The ``large $\phi$'' solution is genuinely three-flavor and 
corresponds to the horizontal depletion zone of the survival probability
$P_{3\nu}^{\rm MSW}$ in Fig.~7. In the first two and last two subplots 
there is no solution in the two-generation limit.

In Fig.~13 we show the fits to single experimental data, in the
same coordinates as in Fig.~12, for three values of $\delta m^2$. 
As in  Fig.~11, the thick, solid lines correspond to iso-$\Delta\chi^2$ 
contours at $95\%$ C.L., and the dotted lines represent iso-signal curves. 
The allowed regions are marked by stars. Notice how, in all cases, the 
solutions in the two-generation limit (lower side, $\phi\rightarrow 0$) 
merge as $\phi$ increases.

\subsection{Solar neutrinos {\em vs\/} accelerator and reactor neutrinos}

In this work we have found and discussed the MSW solutions
to the solar neutrino problem in the parameter space 
$(\delta m^2,\,\tan^2\omega,\,\tan^2\phi)$. In the work \cite{Li95}, 
the results of the established accelerator and reactor neutrino oscillation 
searches were analyzed in  order to constrain the parameter space 
$(m^2,\,\tan^2\psi,\,\tan^2\phi)$. Both works are based on one,
and only one, hypothesis about the neutrino spectrum, namely
$\delta m^2 \ll m^2$ (as in Fig.~1). The angle $\phi$ is common to both
spaces $(\delta m^2,\,\tan^2\omega,\,\tan^2\phi)$ and
$(m^2,\,\tan^2\psi,\,\tan^2\phi)$, as also evidenced in Fig.~5.

We invite the reader to place, side-by-side, Fig.~12 of this work and 
Fig.~10 of \cite{Li95}, in the way suggested by Fig.~5.
These two figures provide us with $12\times12$ combinations
of $(\delta m^2,\,m^2)$. For any given couple of square mass differences 
$(\delta m^2,\,m^2)$, one can find in these two figures the allowed 
regions in the $(\tan^2\omega,\,\tan^2\phi)$ and $(\tan^2\psi,\,\tan^2\phi)$
mixing-mixing planes. We have purposely chosen the same vertical scale 
(same decades in $\log\tan^2\phi$) in both cases. The obvious condition 
that $\tan^2\phi$ (the vertical coordinate) must be the same in both
figures imposes a strong compatibility constraint. This constraint  
eliminates some of the solutions that are individually allowed by solar 
or terrestrial neutrino oscillation searches. For instance, the regions 
allowed by accelerator and reactor data at  very large $\tan^2\phi$ 
(i.e., $\phi\rightarrow\pi/2$) are never allowed by solar neutrino
data, that keep $\tan^2\phi$ below  $\sim1.5$. In other words,  if
$m^2$ is in the range of sensitivity of accelerator and reactor
oscillation searches, then the comparison of terrestrial and solar data 
requires relatively small values of $\phi$. Another example is that the 
MSW-allowed values $\delta m^2\simeq1.5\times10^{-4}$ eV$^2$
and $\delta m^2\simeq(2.2$--$3.2)\times10^{-6}$ eV$^2$ are not compatible
with accelerator and reactor data for $m^2\gtrsim10^{-2}$ eV$^2$.
One can explore various other  (in)compatibility situations for the 
$12\times12$ cases of $(\delta m^2,\,m^2)$ that can be extracted from 
the figures discussed above.

We recall that the only hypothesis adopted in our analysis is 
$\delta m^2 \ll m^2$. Virtually, all see-saw models of neutrino mass 
generation satisfy this condition. Model builders can thus check if their 
favorite values of $(\delta m^2,\,m^2,\,\omega,\,\psi,\,\phi)$ fall in
the allowed regions determined in this work and in \cite{Li95}.
Here we discuss briefly the ``threefold maximal mixing'' model, recently
investigated in \cite{Max2}.

The maximal mixing scenario is realized in the centers of the triangles
in Fig.~5, that is for $U^2_{ei}=U^2_{\alpha3}=1/3$ or, equivalently,
for $(\tan^2\omega,\,\tan^2\psi,\,\tan^2\phi)=(1,\,1,\,1/2)$.
Figure~9 shows that the combination $(\tan^2\omega,\,\tan^2\phi)=(1,\,1/2)$
is excluded at more than $99\%$ C.L. by solar $\nu$ data
(within the MSW approach). The maximal mixing scenario could be 
reconciled with solar $\nu$ data only by dropping the Homestake measurement, 
that keeps $\omega$ below $\pi/4$. From Fig.~10 in  \cite{Li95}, it follows 
that accelerator and reactor data exclude the maximal 
mixing combination $(\tan^2\psi,\,\tan^2\phi)=(1,\,1/2)$
for all values of $m^2$ in their present sensitivity range  
($m^2\gtrsim 10^{-2}$ eV$^2$). Thus, the maximal mixing is not supported 
by all solar neutrino data, and is (trivially) compatible with accelerator 
and reactor experiments only below their present $m^2$-sensitivity limits.

\section{SUMMARY  AND  CONCLUSIONS}

We have discussed the three-flavor MSW mechanism, in the hypothesis
$\delta m^2\ll m^2$ and for mixing angles $\omega$ and $\phi$ in
their maximal range $[0,\,\pi/2]$. New useful graphical representations 
of the parameter space (Figs.\ 4--6) have been introduced.
A complete analysis of present data in the mass-mixing space 
$(\delta m^2,\,\omega,\,\phi)$ has been performed, and the solutions
have been thoroughly discussed. The data favor the two-generation limit
$(\phi=0^{\circ})$, but a wide range of three-flavor solutions are  allowed 
up to $\phi\simeq50^{\circ}$. The main results are summarized in Figs.~9 
and 12.

The $\nu_e$ oscillation probability has been studied as a function
of the mass and mixing parameters (Figs.\ 7 and 8). In the allowed 
regions at large $\phi$, the three-flavor probability varies more slowly 
with energy than in the two-generation case. Thus, it will be difficult 
to study the large-$\phi$ solutions in the future solar neutrino experiments
that are meant to probe the energy dependence of the solar neutrino flux.

It has been shown that the solar neutrino results are linked to those
obtained in an analysis of accelerator and reactor neutrino oscillation 
searches \cite{Li95}, where the mass-mixing space  $(m^2,\,\psi,\,\phi)$ 
has been thoroughly explored (Fig.~5). This link can be used 
to further constrain the allowed ranges of the parameters. In particular, 
for $m^2\gtrsim 10^{-2}$ eV$^2$, the angle $\phi$ is constrained to 
relatively small values. The threefold maximal mixing scenario 
\cite{Max2} is disfavored by solar neutrino data (in the whole explored 
range of $\delta m^2$)  as well as by accelerator and reactor experiments
for  $m^2\gtrsim10^{-2}$.

The present study and the work in \cite{Li95} represent a detailed and 
accurate analysis of solar, accelerator, and  reactor neutrino oscillations
in three-flavors. The analysis has made it possible to obtain
significant bounds in the mass-mixing parameter space 
$(\delta m^2,\,m^2,\,\omega,\,\psi,\,\phi)$.

When this work was completed, we become aware of a recent three-flavor
analysis of solar neutrino data performed by Narayan {\em et al.\/} in
\cite{Na96}. We agree  only qualitatively on their results
about the MSW solutions. We do not agree on the precise shape of the 
allowed regions, and do not confirm their claim that the solutions are 
dramatically enlarged when the  confidence level is moderately increased 
(from $1\sigma$ to $1.6\sigma$ in their work). The information contained 
in \cite{Na96} is not detailed enough to trace  the differences.

\acknowledgments

We are grateful to J.~N.\ Bahcall, P.~I.\ Krastev, and S.~T.\ Petcov for 
useful discussions and suggestions. 
One of us (E.L.) thanks the organizers of the ``1995 Santa Fe Workshop on 
Massive Neutrinos'' (Santa Fe, New Mexico, August 1995), where this work
was initiated, for having provided a stimulating and friendly environment.
The work of E.L.\ was supported by 
INFN and by a Hansmann fellowship at the Institute for Advanced Study.
This work was performed under the auspices of the Theoretical Astroparticle
Network (contract No.\ CHRX-CT93-0120 of the Dir.\ Gen.\ XII of the E.E.C.).


\begin{table}
\caption{	Neutrino rates observed in the four operating solar 
		neutrino experiments \protect\cite{Da94,GALL,SAGE,Kami},
 		and corresponding theoretical predictions of the Standard 
		Solar Model (SSM) by Bahcall and Pinsonneault 
		\protect\cite{Ba95}.}
\begin{tabular}{lcccc}
Experiment & Ref.                     & Measured rate $\pm$(stat)$\pm$(syst) 
           & Units                    & SSM rate \protect\cite{Ba95}   \\     
\hline
Homestake  & \protect\cite{Da94}      &$  2.55 \pm 0.17 \pm0.18       $
           & SNU                      &$  9.3  ^{+1.2}_{-1.4}         $\\
GALLEX     & \protect\cite{GALL}      &$  77.1 \pm 8.5^{+4.4}_{-5.4}  $
           & SNU                      &$  137  ^{+8}_{-7}             $\\
SAGE       & \protect\cite{SAGE}      &$  69   \pm 11^{+5}_{-7}       $
           & SNU                      &$  137  ^{+8}_{-7}             $\\
Kamiokande & \protect\cite{Kami}      &$  2.89 \pm 0.22 \pm0.35       $
           & $10^6$ cm$^{-2}$s$^{-1}$ &$  6.62 ^{+0.93}_{-1.13}       $
\end{tabular}
\end{table}

\newcommand{\InsertFigure}[2]{\newpage\begin{center}\mbox{%
\epsfig{bbllx=1.4truecm,bblly=1.3truecm,bburx=19.5truecm,bbury=26.5truecm,%
height=21.truecm,figure=#1}}\end{center}\vspace*{-1.85truecm}%
\parbox[t]{\hsize}{\small\baselineskip=0.5truecm\hskip0.5truecm #2}}
\InsertFigure{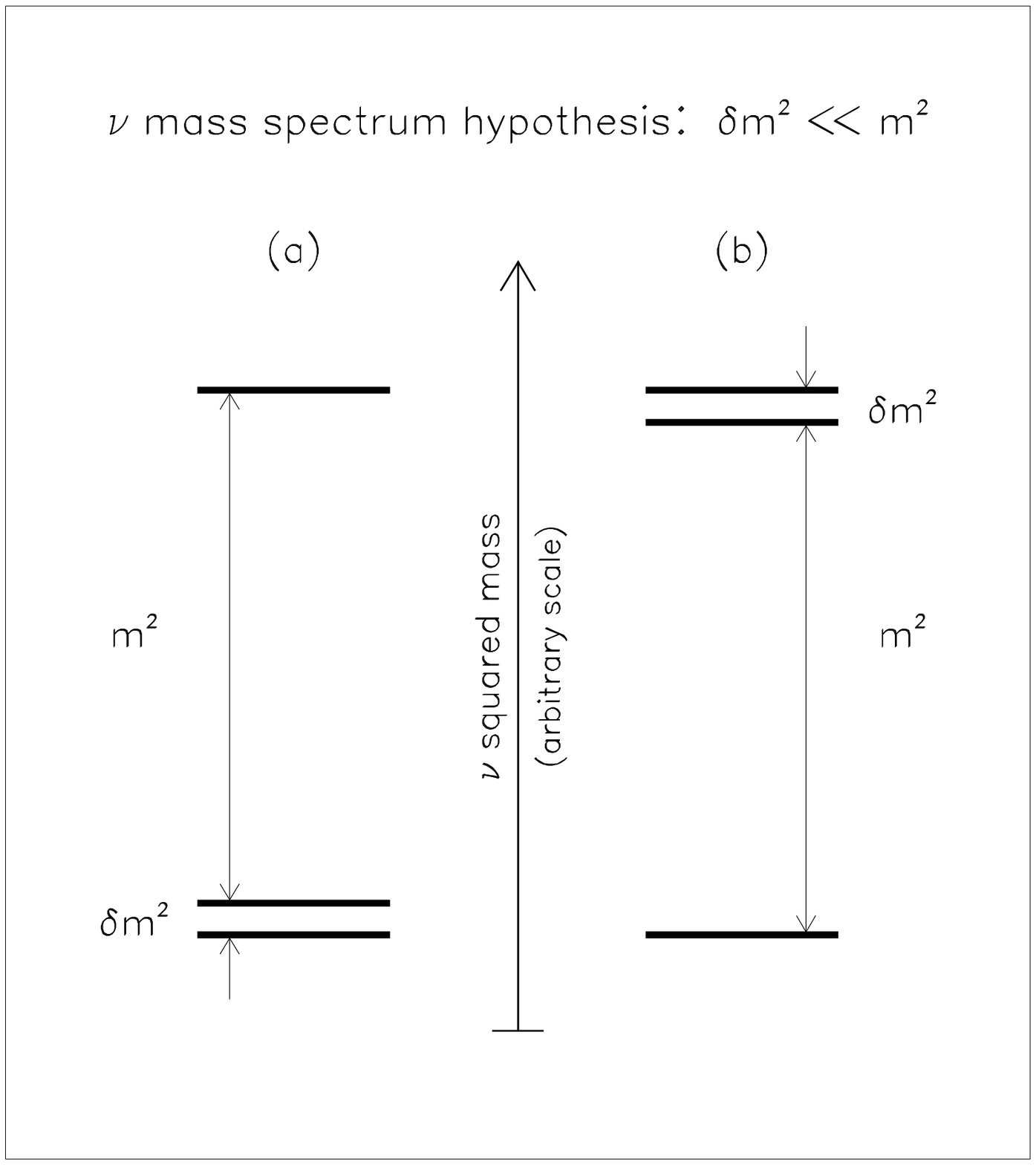}%
{FIG.~1. 	The neutrino mass spectrum assumed in this work. The 
		hypothesis $\delta m^2 \ll m^2$ is independent of the 
		zero of the absolute mass scale. Solar neutrinos do not 
		distinguish the two cases (a) and (b)  at zeroth order 
		in $\delta m^2/m^2$.}
\InsertFigure{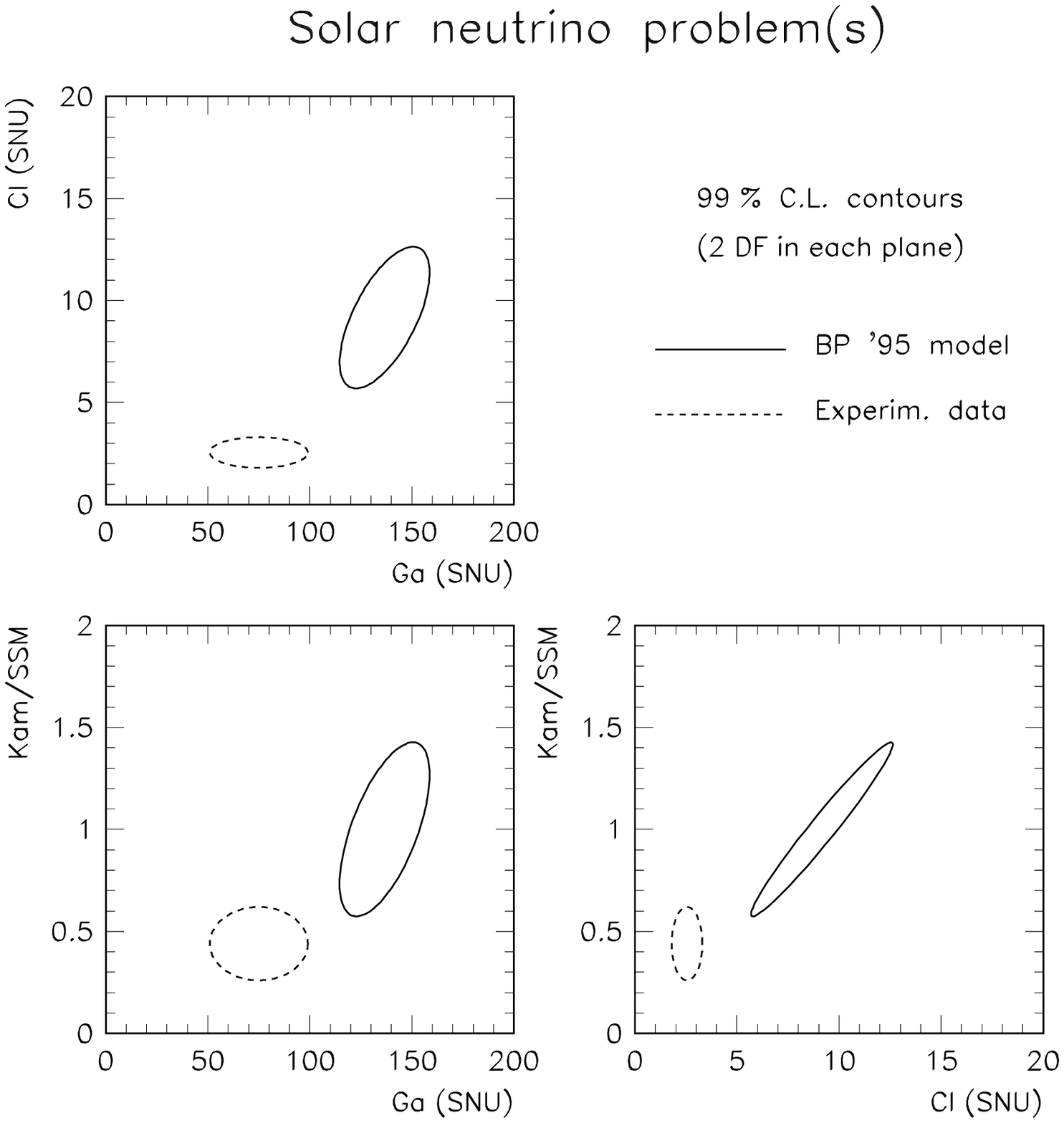}%
{FIG.~2.	The regions allowed at $99\%$ C.L. by the present solar 
		neutrino data (dashed lines) and by the standard solar 
		model (SSM) of  Bahcall-Pinsonneault \protect\cite{Ba95} 
		(solid lines). The coordinates are the chlorine (Cl), 
		gallium (Ga), and water-Cherenkov (normalized Kam/SSM) 
		signals.}
\InsertFigure{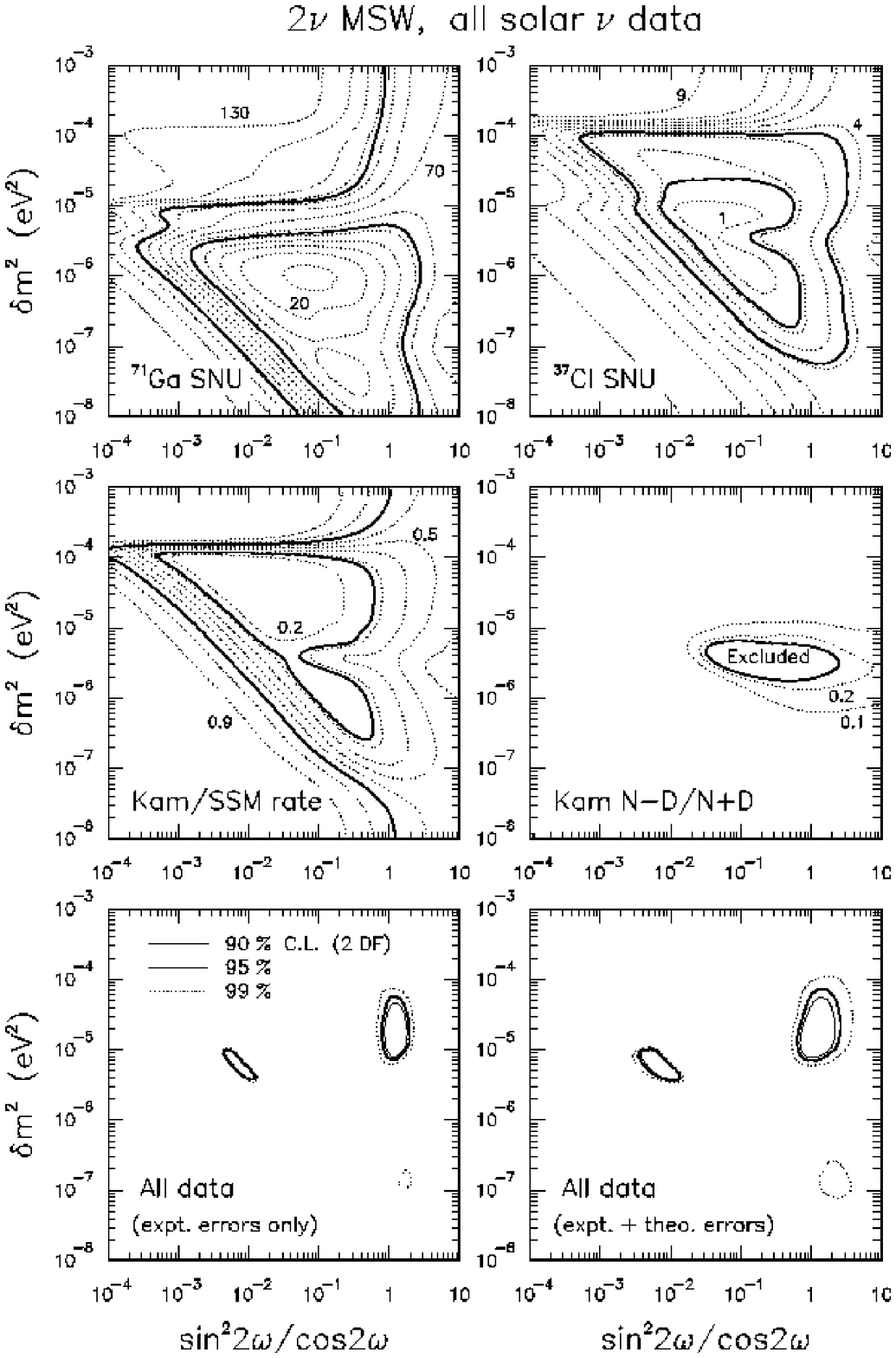}%
{FIG.~3.	Two-generation MSW solutions to the solar neutrino problem.
		In the upper four panels (fits to single data), the solid 
		lines define the regions allowed at $95\%$ C.L.\ for two 
		degrees of freedom (including theoretical errors), and the 
		dotted lines represent iso-signal contours. In the lower two 
		panels (fits to all data), the MSW solutions are shown at 
		$90\%$, $95\%$, and $99\%$ C.L., without and with theoretical 
		errors.} 
\InsertFigure{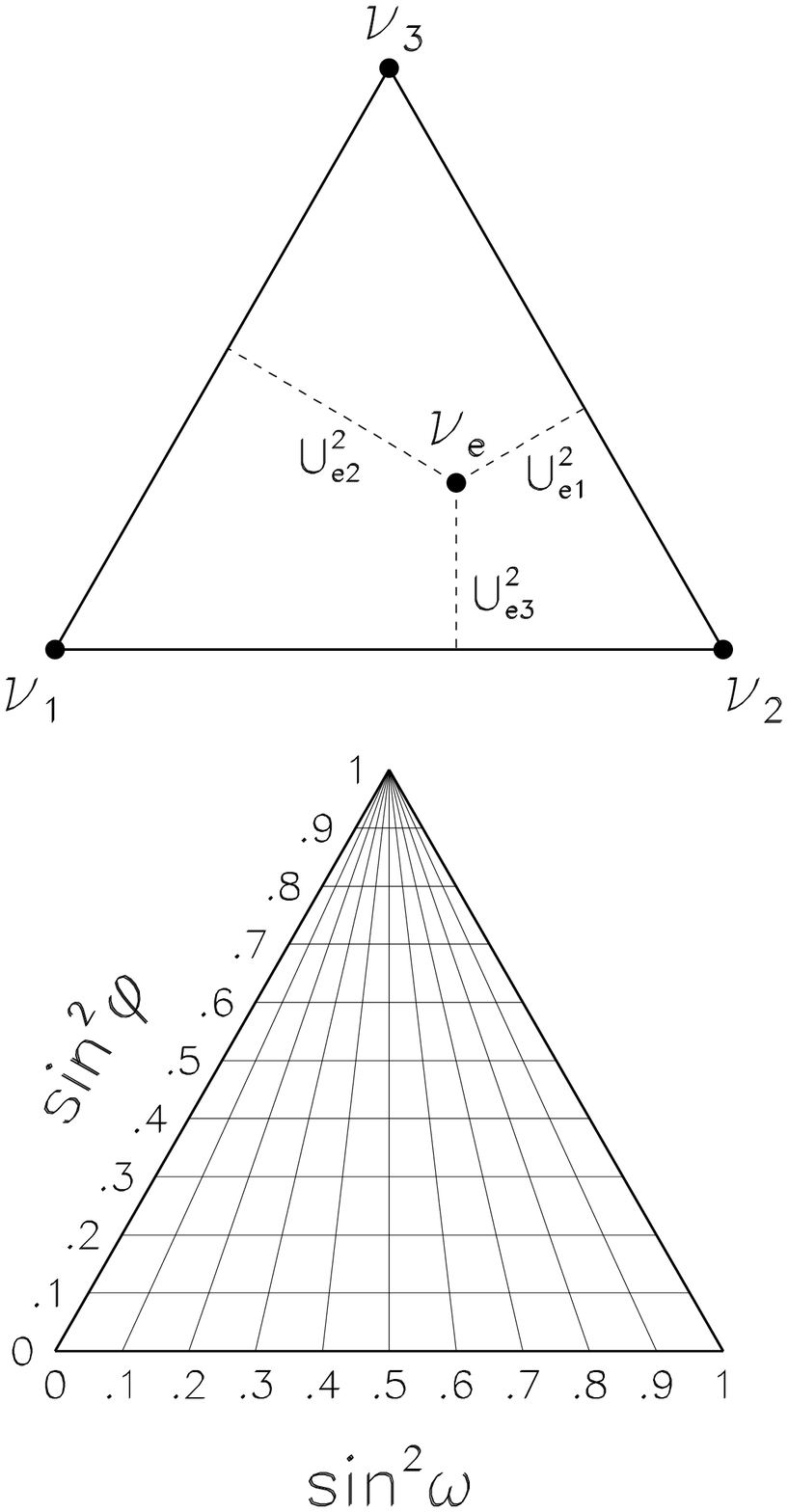}%
{FIG.~4.	Triangular plot representing  the neutrino state $\nu_e$ in 
		terms of its massive components $\nu_1,\,\nu_2,\,\nu_3$. For 
		a triangle of unit height, the unitarity relation 
		$U^2_{e1}+U^2_{e2}+U^3_{e3}=1$ is enforced by identifying the
		$U^2_{ei}$ with the three distances of $\nu_e$ from the sides
		(dashed lines). The usual two-generation oscillation limit
		is reached for $\nu_e$ onto the lower side joining $\nu_1$ 
		to $\nu_2$. Iso-lines of $\sin^2\omega$ and $\sin^2\phi$ 
		(parameterizing the elements $U^2_{ei}$) are charted in the 
		lower plot.}
\InsertFigure{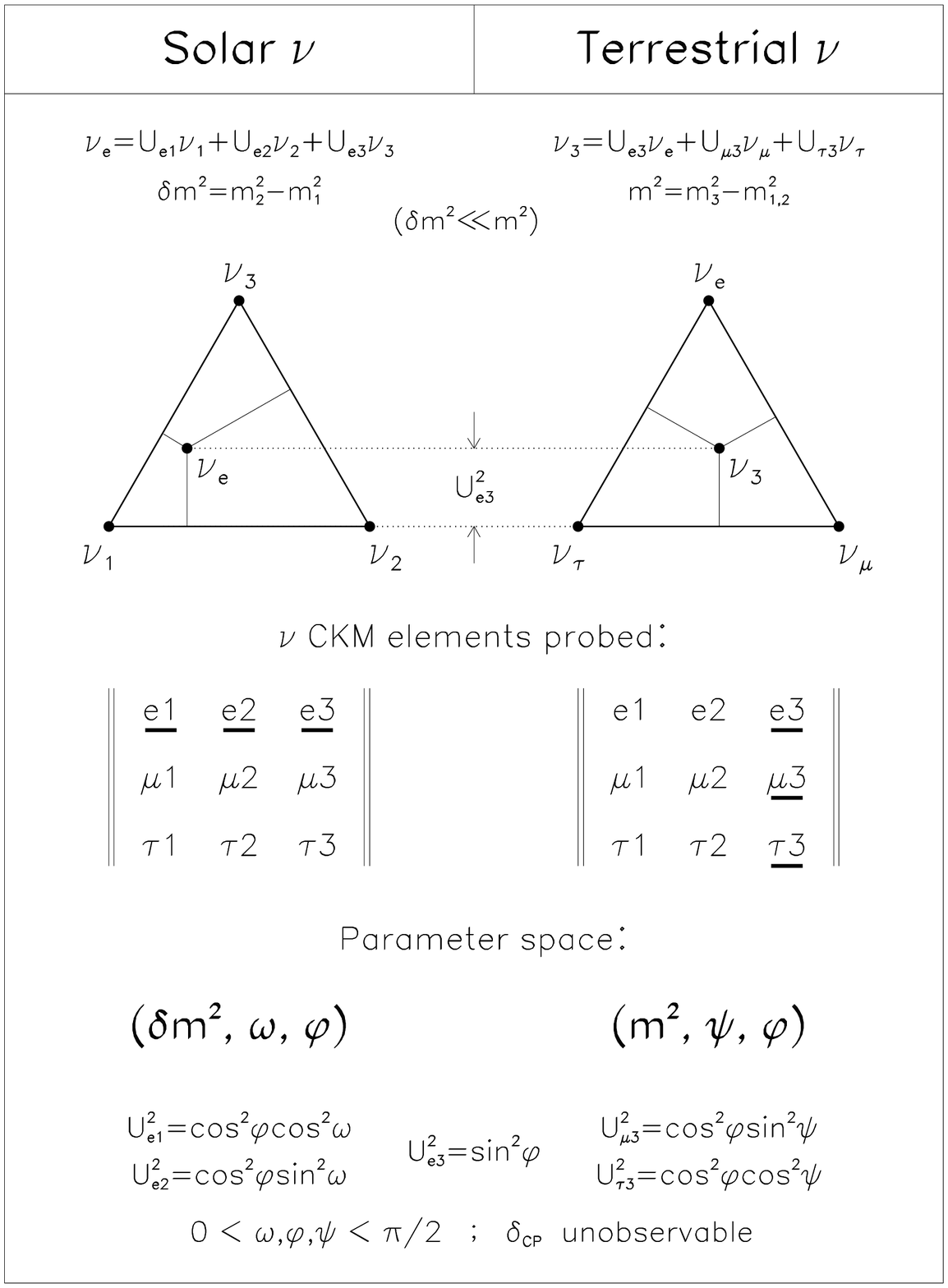}%
{FIG.~5.	A synoptic presentation of the parameter space for
		solar and terrestrial (accelerator, reactor, atmospheric)
		neutrino oscillation searches. Solar neutrinos probe the 
		small square mass difference, $\delta m^2$, and the mixing 
		matrix elements $U^2_{ei}$. Terrestrial neutrinos probe the 
		large square mass difference, $m^2$, and the mixing matrix 
		elements $U^2_{\alpha3}$. Notice that $U^2_{e3}$ is probed 
		in both cases, whilst the CP-violating phase $\delta$ is 
		unobservable. The expansion of $U^2_{\alpha i}$ in terms of 
		the mixing angles $(\omega,\,\phi,\,\psi)$ is shown. The 
		reader is referred to the text and to
		\protect\cite{Li95} for further details.} 
\InsertFigure{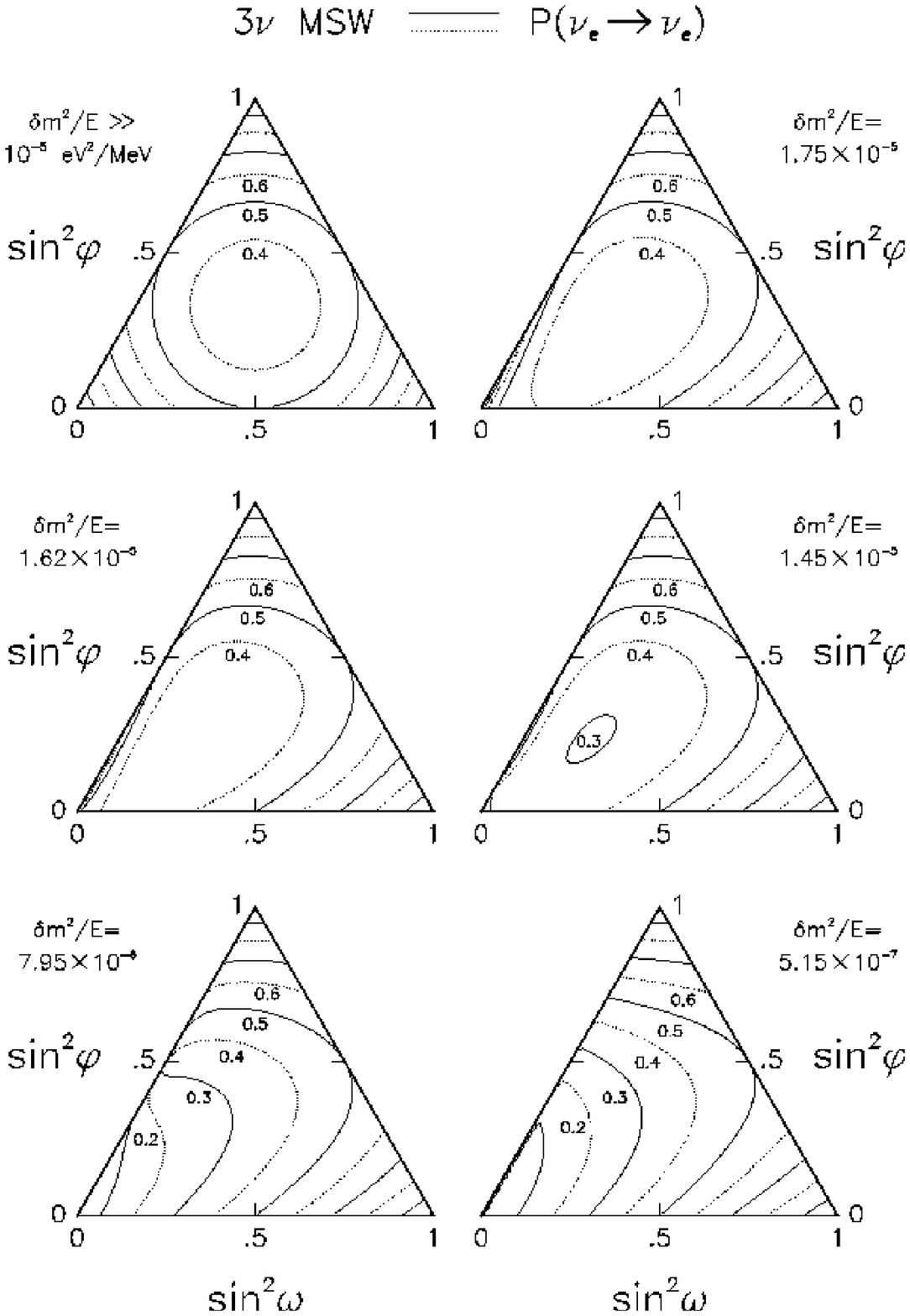}%
{FIG.~6. 	Iso-lines of survival probability $P(\nu_e\rightarrow\nu_e)$
		in the triangular plot defined in Fig.~4, for six 
		representative values of $\delta m^2/E_\nu$. The first 
		subplot represents the averaged vacuum oscillation case.
		In the other subplots, the MSW effect is active and produces 
		an increasing deformation of the curves as $\nu_e$ approaches 
		$\nu_3$ (the left lower corner).  For 
		$\omega=0,\,\pi/4,\,\pi/2$,  however, the probability remains 
		equal to the vacuum value, as discussed in the text. 
		(The behavior at $\omega\simeq 0$ cannot be graphically 
		resolved in the last two subplots.)}
\InsertFigure{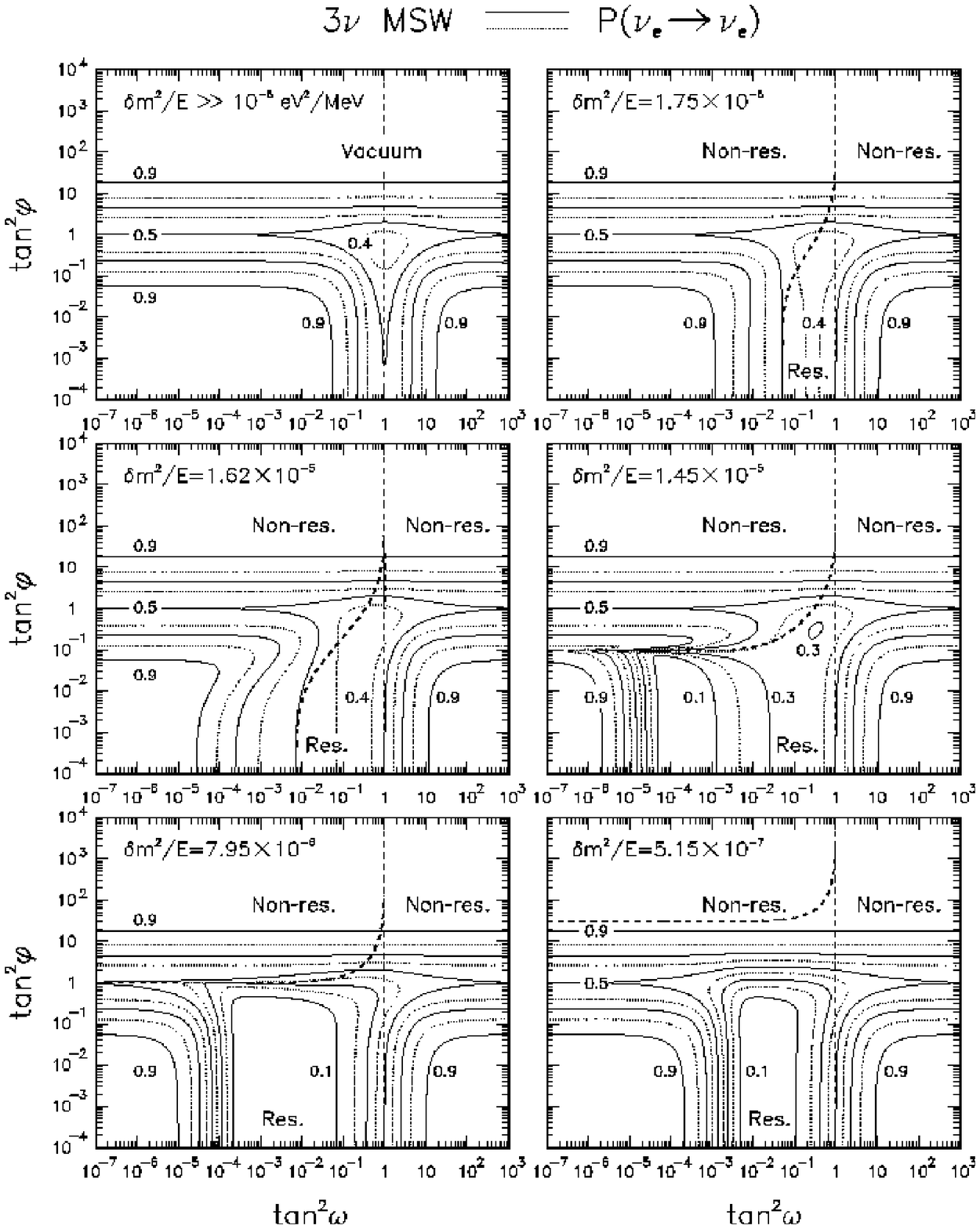}%
{FIG.~7.	As in Fig.~6, but in the bilogarithmic 
		$(\tan^2\omega,\,\tan^2\phi)$ representation. Notice the 
		deformation of the iso-probability contours due to the 
		$\triangle \rightarrow \Box$ mapping. In this representation
		the $\nu_1,\,\nu_2,\,\nu_3$ mass eigenstates are reached 
		respectively in the limits:
		$\tan^2\phi\rightarrow0$ and $\tan^2\omega\rightarrow0$;
		$\tan^2\phi\rightarrow0$ and $\tan^2\omega\rightarrow\infty$;
		$\tan^2\phi\rightarrow\infty$  at any $\tan^2\omega$. 
		The thick, dashed lines represents the curves for maximal 
		$\omega$-mixing in vacuum and matter: $\omega=\pi/4$ 
		(vertical) and  $\omega^0_m=\pi/4$ (curve on the left). 
		These two curves separate the zones where the neutrino
		propagation is non-resonant from those where it is resonant.} 
\InsertFigure{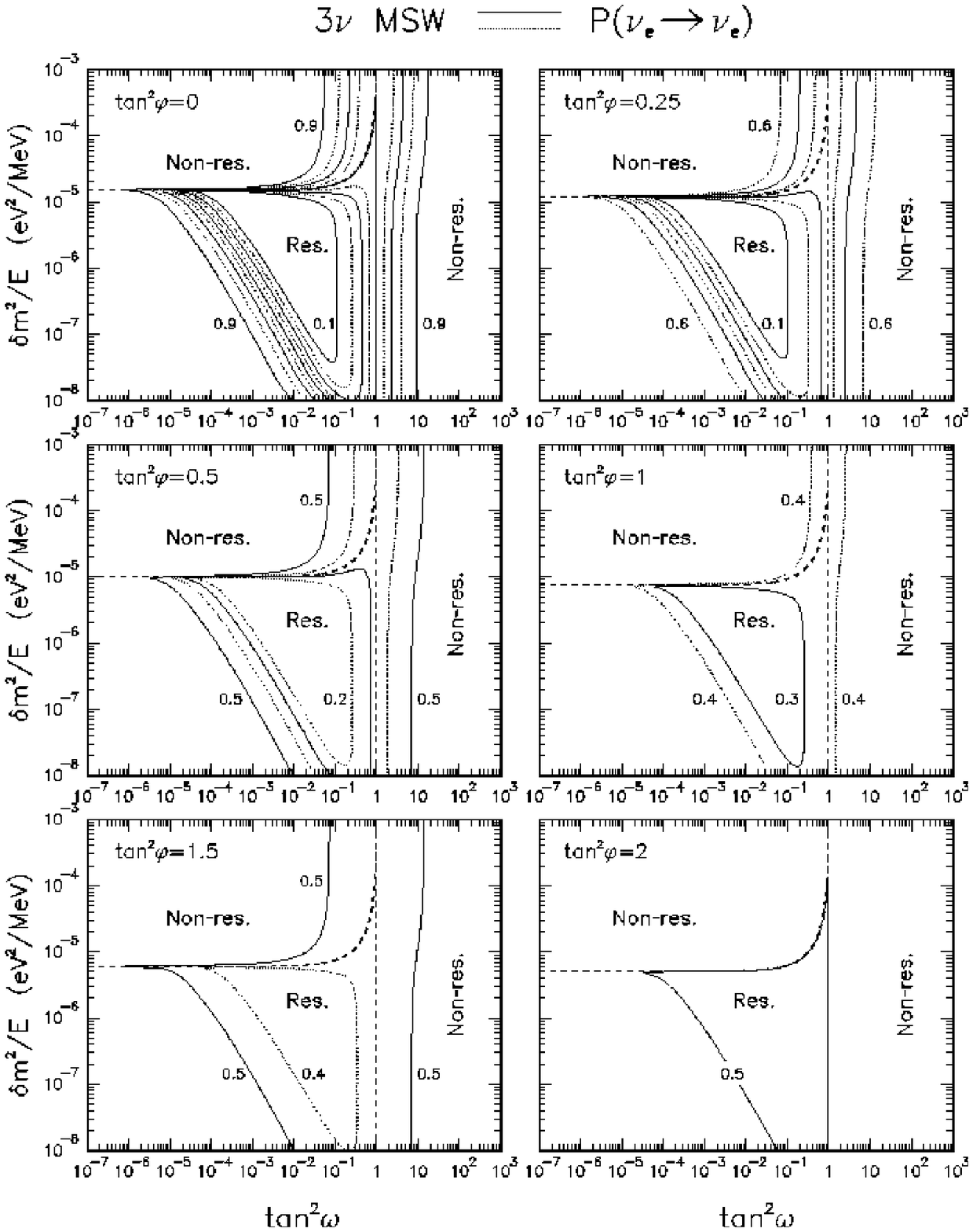}%
{FIG.~8.	Iso-lines of survival probability $P(\nu_e\rightarrow\nu_e)$
		in the bilogarithmic $(\tan^2\omega,\,\delta m^2/E_\nu)$ 
		plane, for representative values of $\tan^2\phi$. The thick, 
		dashed curves are defined  as in Fig.~7. The resonant 
		region has the characteristic triangular shape.
		The variations of $P(\nu_e\rightarrow\nu_e)$ in the plane 
		decrease for increasing $\phi$.} 
\InsertFigure{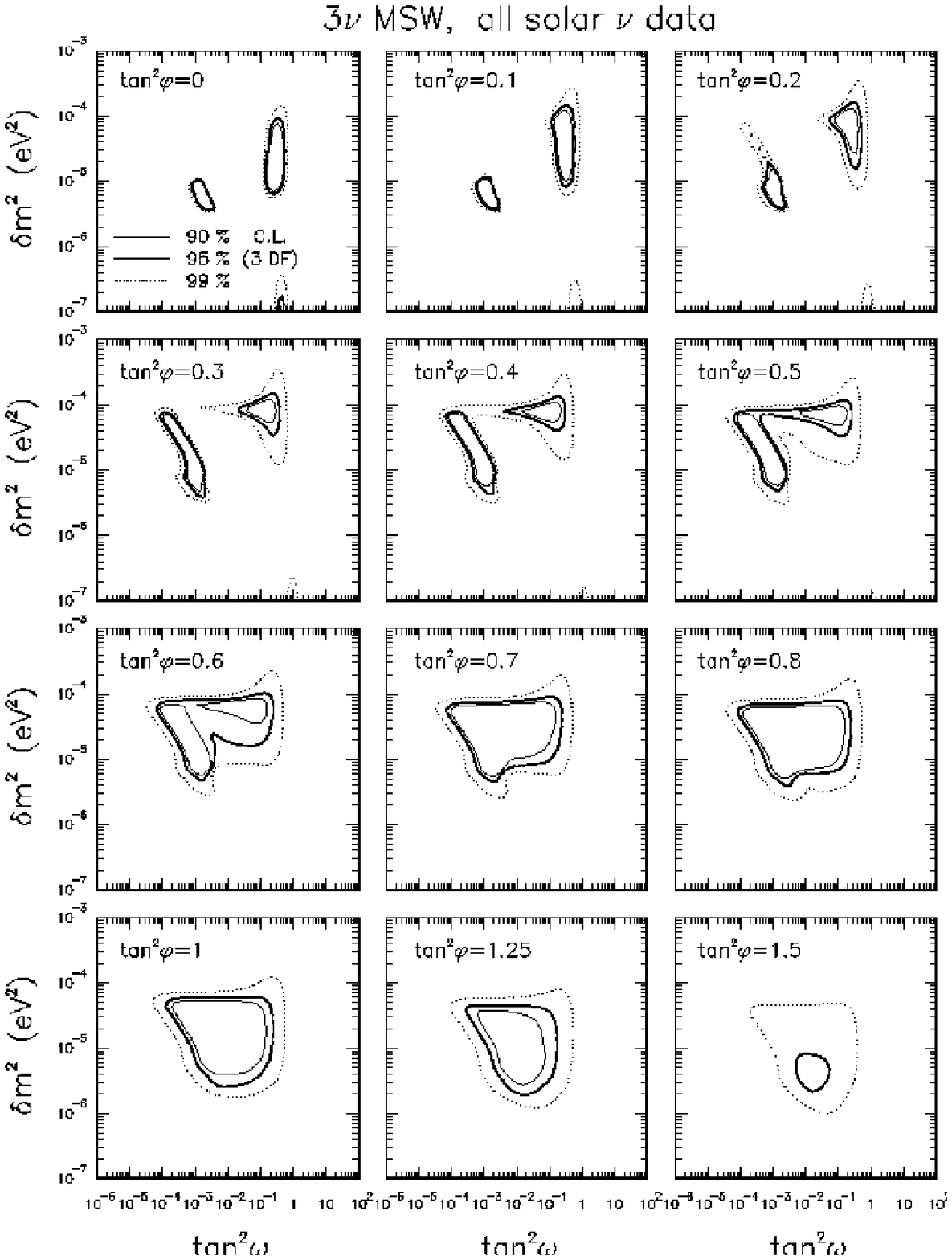}%
{FIG.~9.	Three-flavor MSW analysis of all solar neutrino
		data. The regions allowed at $90\%$, $95\%$, $99\%$ C.L.\ 
		in the space $(\delta m^2,\,\tan^2\omega,\,\tan^2\phi)$ 
		are shown in planar $(\delta m^2,\,\tan^2\omega)$ sections 
		at twelve representative values of $\tan^2\phi$
		ranging from 0 to 1.5. The two-generation limit is recovered
		at $\tan^2\phi=0$ (first subplot). The two separate 
		solutions at small $\phi$ merge in one single solution for 
		increasing $\phi$. See the text for details.}
\InsertFigure{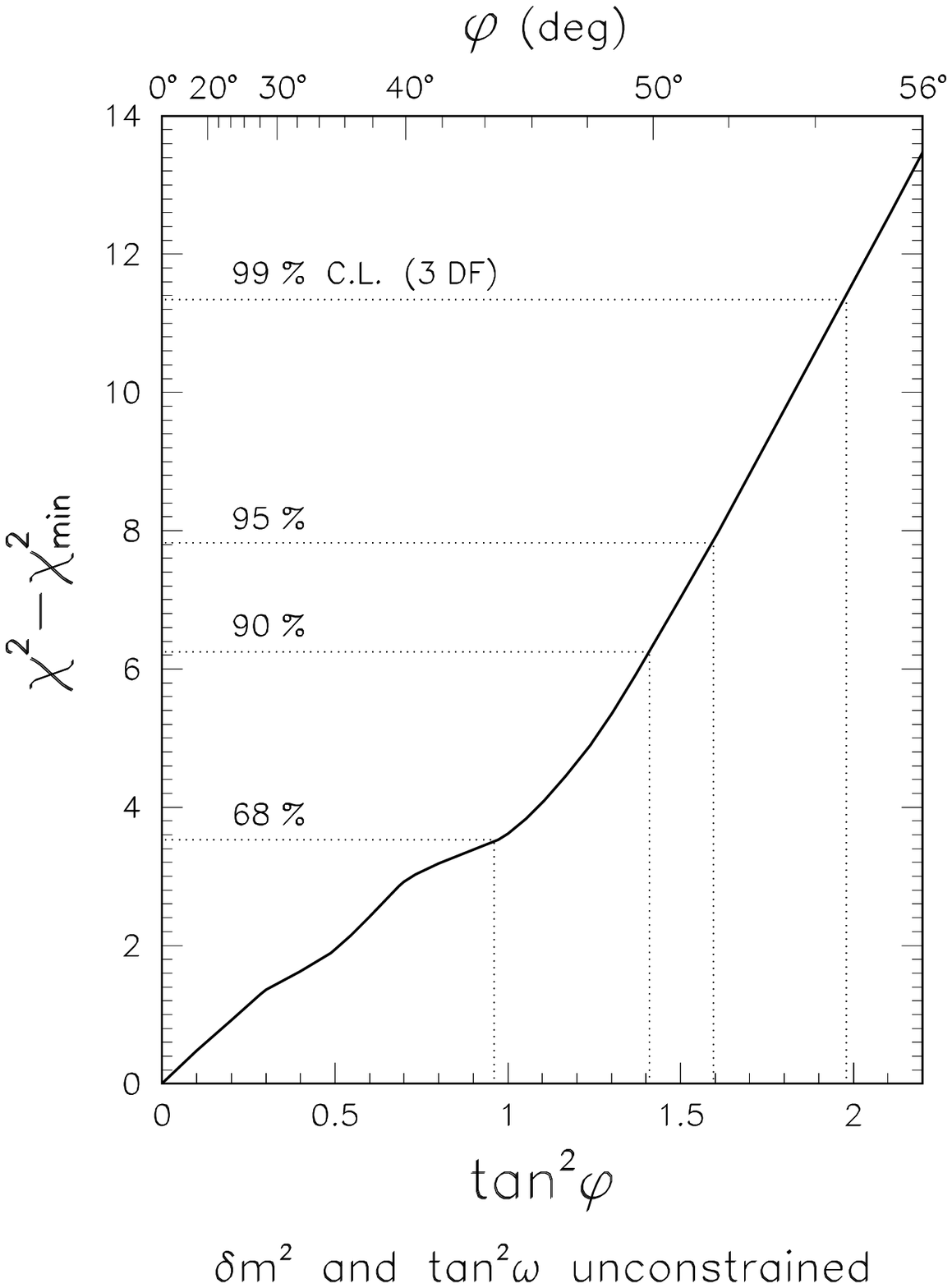}%
{FIG.~10.	Values of $\Delta \chi^2$ as a function of  $\tan^2\phi$, 
		for $\delta m^2$ and $\tan^2\omega$ unconstrained. The 
		value $\phi=0$, corresponding to the two-generation limit,
		is preferred, but values as high as $\tan^2\phi\simeq1.4$
		$(\phi\simeq50^\circ)$ are still allowed at $90\%$ C.L.} 
\InsertFigure{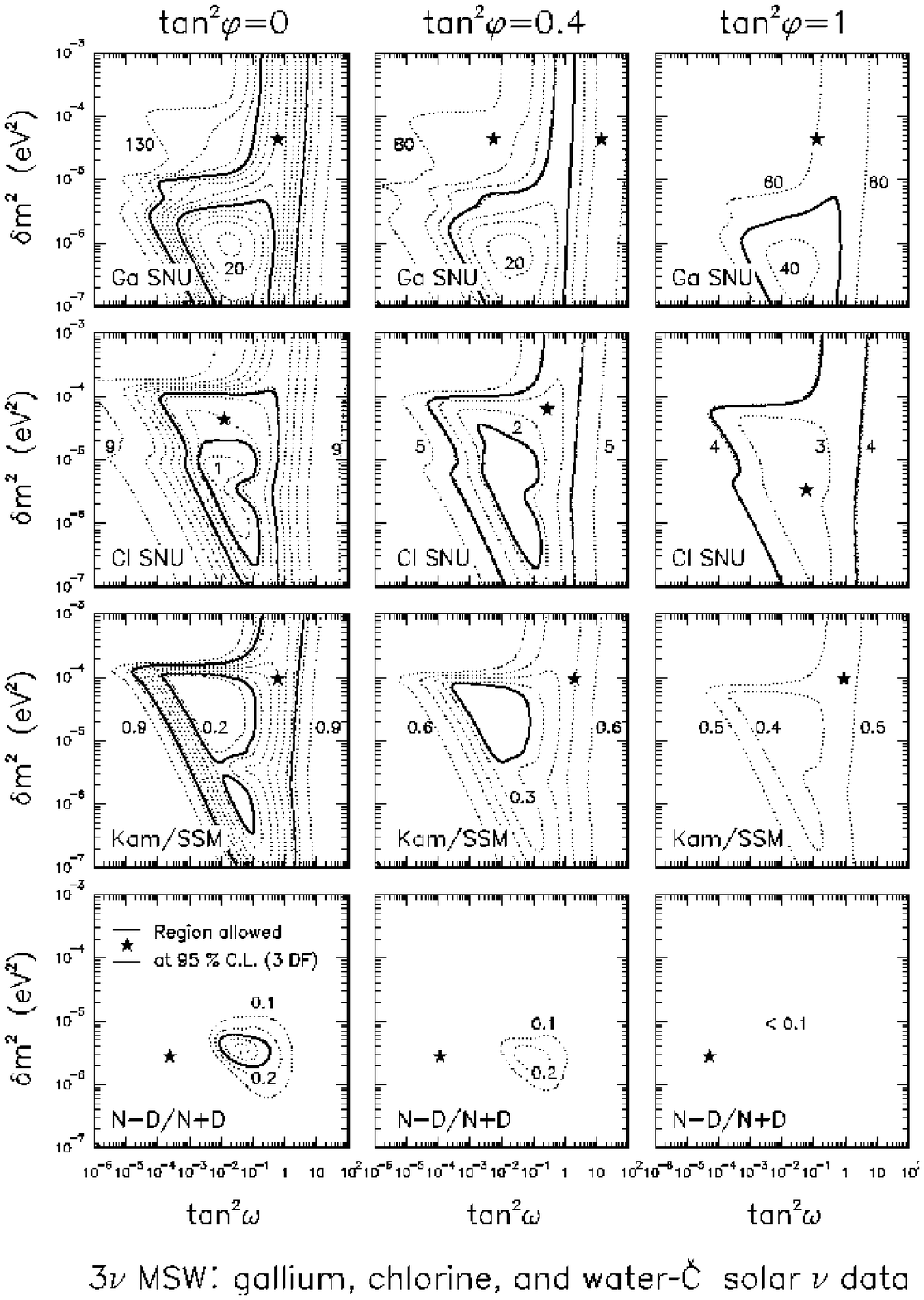}%
{FIG.~11.	Three-flavor MSW analysis of individual neutrino
		data in the same coordinates as in Fig.~9, for three 
		representative values of $\tan^2\phi$. The solid lines 
		correspond to $95\%$ C.L.\ contours. The allowed regions 
		are marked by stars. Dotted lines represent iso-signal
		contours.} 
\InsertFigure{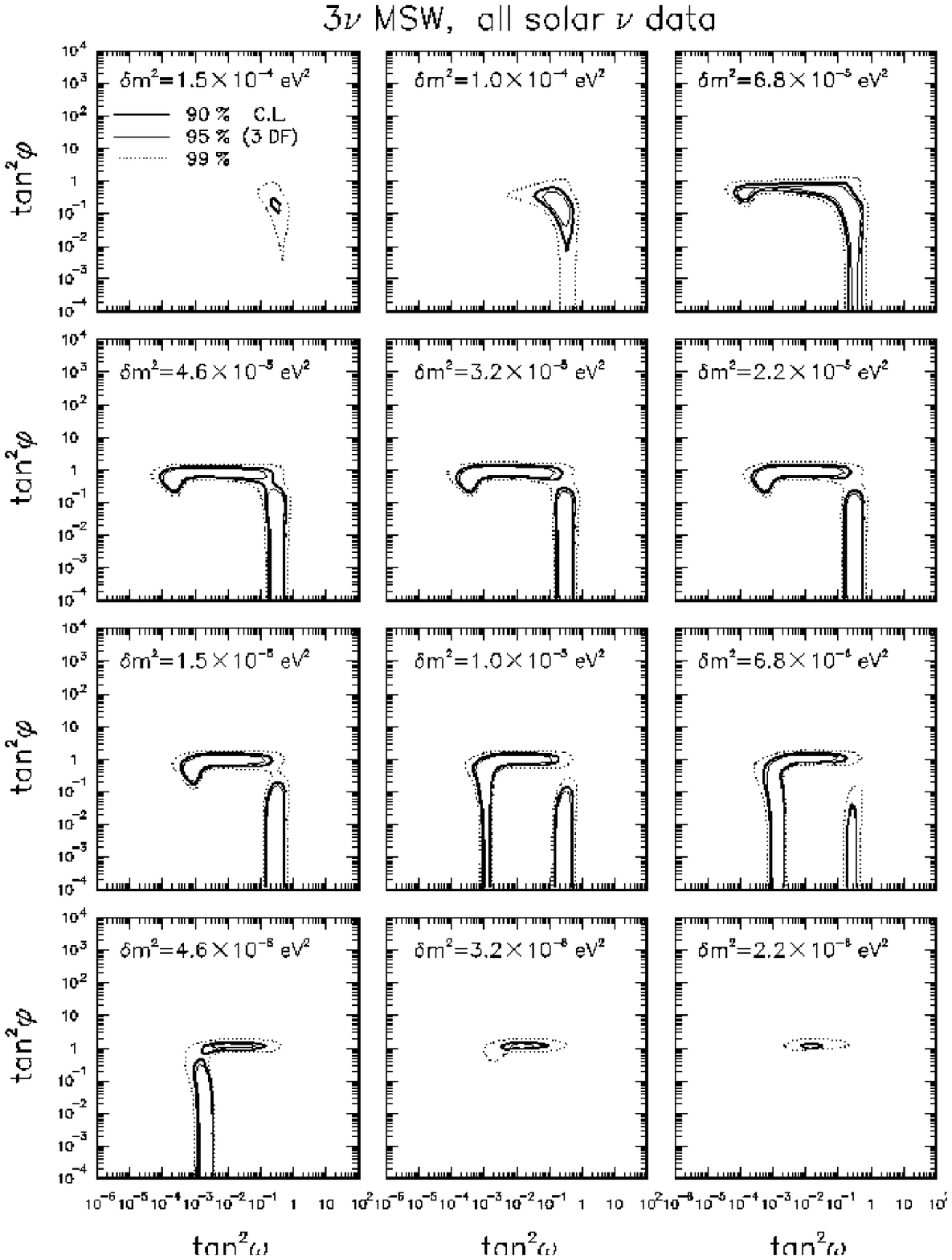}%
{FIG.~12.	Results of the three-flavor MSW analysis of all solar 
		neutrino data. The regions allowed at $90\%$, $95\%$, 
		$99\%$ C.L.\ in the space 
		$(\delta m^2,\,\tan^2\omega,\,\tan^2\phi)$ are shown in 
		planar $(\tan^2\omega,\,\tan^2\phi)$ sections at twelve 
		representative values of $\delta m^2$ ranging from
		$1.5\times10^{-4}$ to $2.2\times 10^{-6}$ eV$^2$.
		The two-generation limit is recovered for 
		$\tan^2\phi\rightarrow0$ (lower side of each subplot).  
		Notice the appearance of solutions disconnected from the 
		two-generation limit. See the text for details.}	
\InsertFigure{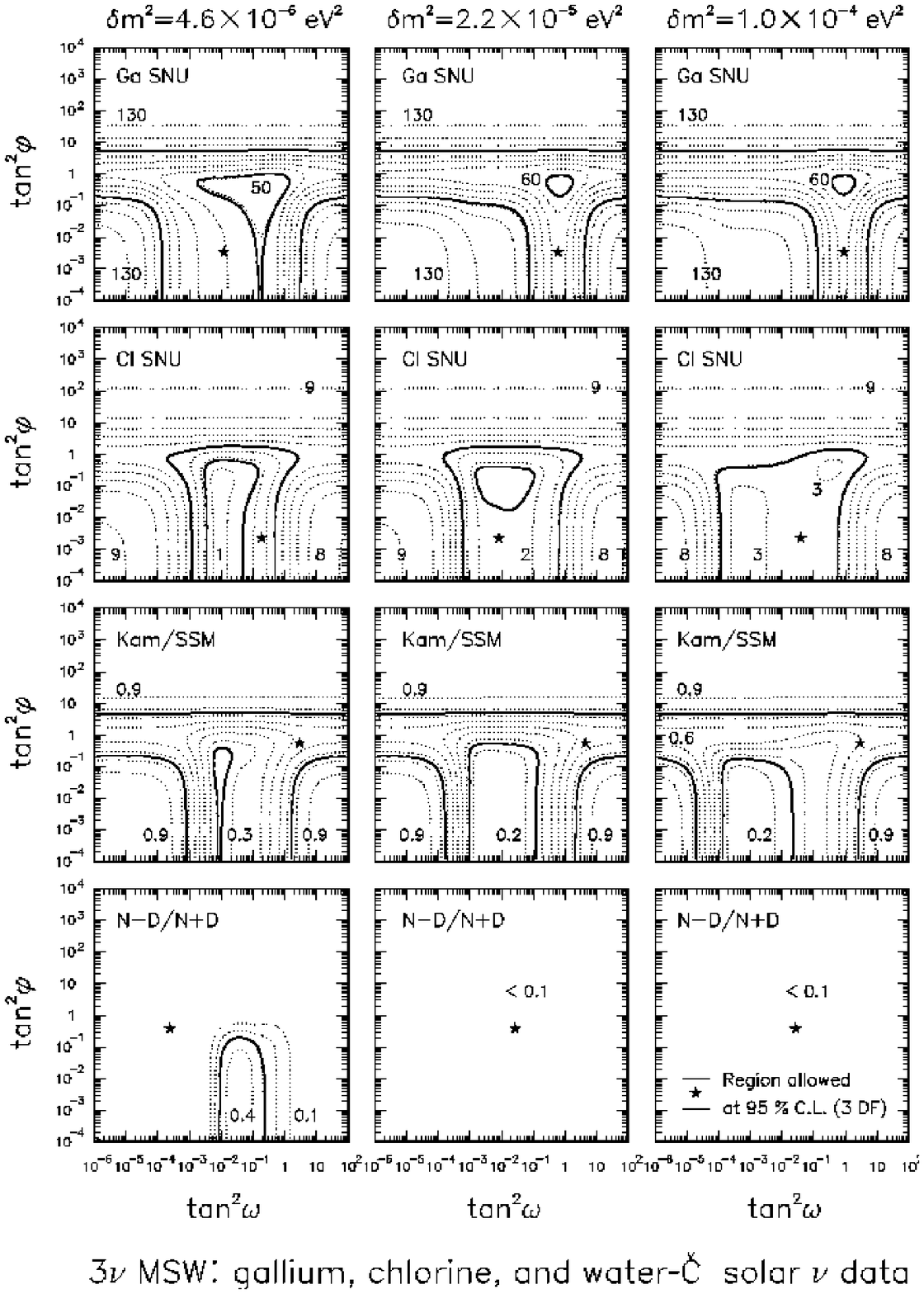}%
{FIG.~13. 	Results of the three-flavor MSW analysis of individual 
		neutrino data in the same coordinates as in Fig.~12 for 
		three representative values of $\delta m^2$. The solid 
		lines correspond to $95\%$ C.L.\ contours. The allowed 
		regions are marked by stars. Dotted lines represent 
		iso-signal contours.}

\end{document}